\documentclass[12pt]{article}

\usepackage{amsmath}
\usepackage{amsfonts,amssymb,amsthm,overpic}
\makeatletter
\newcommand{\diam}{\mathop{\operator@font diam}}
\usepackage{setspace}
\usepackage{multicol}
\usepackage{multirow}
\usepackage[compact]{titlesec}

\usepackage{epsfig}

\newtheorem{theorem}{Theorem}[section]

\newtheorem{corollary}{Corollary}[section]

\newcommand{\cT}{\mathcal{T}}

\begin{document}

\title{\Huge{\textsc{Spacetimes as topological spaces, and the need to take methods of general topology more seriously}}}

\author{Kyriakos Papadopoulos$^1$ and Fabio Scardigli$^{2, 3}$ \\
\small{1. Department of Mathematics, Kuwait University, PO Box 5969, Safat 13060, Kuwait} \\
\small{2. Dipartimento di Matematica, Politecnico di Milano, Piazza Leonardo da Vinci 32, 20133 Milano, Italy} \\
\small{3. Institute-Lorentz for Theoretical Physics, Leiden University, P.O.~Box 9506, Leiden, The Netherlands}}

\date{}

\maketitle

\begin{abstract}
Why is the manifold topology in a spacetime taken for granted? Why
do we prefer to use Riemann open balls as basic-open sets, while there also exists a Lorentz metric? Which
topology is a best candidate for a spacetime; a topology sufficient
for the description of spacetime singularities or a topology which
incorporates the causal structure? Or both? Is it more preferable to
consider a topology with as many physical
properties as possible, whose description might be complicated and
counterintuitive, or
a topology which can be described via a countable basis but misses some
important information? These are just a few from the questions
that we ask in this Chapter, which serves as a critical review of the terrain
and contains a survey with remarks, corrections and open questions.
\end{abstract}

{\bf AMSC:} 83–XX, 83F05, 85A40, 54–XX    \\
{\bf keywords:} Zeeman - G\"obel topologies, topologising a spacetime,
spacetime singularities, causal topologies, manifold topology

\section{Introduction.}

\subsection{The Manifold Topology vs. Finer or Incomparable Topologies.}

In \cite{Heathcote}, the author supports that the manifold topology in a curved spacetime
is the best possible and most natural choice, against the class of topologies that was
suggested by Zeeman and G\"{o}bel (see \cite{Zeeman1} and \cite{gobel}, respectively). His
main focus lies on topologies finer than the manifold one, but there is a misjudgemet here: there are topologies in the class $\mathfrak{Z}$ of Zeeman-G\"{o}bel topologies (as we shall see in paragraph 4), that are neither
finer nor equal nor coarser than the manifold topology.

Instead of asking whether we need a finer topology for a sufficient mathematical
description of a spacetime, we bring the topologisation question into a different level;
why should one prefer a topology which describes spacetime singularities against a topology which hides singularities but incorporates the causal structure
of the spacetime? As we shall see, the singularity theorems were proven under the frame of the
manifold topology, while there are topologies in the class $\mathfrak{Z}$ where
the Limit Curve Theorem (LCT for abbreviation) fails to hold, and thus sufficient conditions
for the formation of singularities as we understand them in the presence of Riemannian basic-open balls
fail as well.

Zeeman's main arguments against the Euclidean $\mathbb{R}^4$ topology for Minkowski spacetime $M$ (extended by G\"{o}bel for curved spacetimes)
can be summarised as follows:

\begin{enumerate}
\item The $4$-dimensional Euclidean topology is locally homogeneous, whereas $M$ is
not; every point has associated with it a light cone, separating space vectors from time
vectors.

\item The group of all homeomorphisms of $4$-dimensional Euclidean space is vast, and
of no physical significance.
\end{enumerate}

Heathcote's antilogue belongs to (sic) a realist view of spacetime topology as against
the instrumentalist position. A realist point of view divides the space intro structural
levels, such as metric tensor field, affine connection, conformal structure, differentiable
manifold and topology. Heathcote highlights that the manifold topology is present as long
as the structure of manifold is present, and there are two ``untenable'' possibilities for
a replacement of the manifold topology, in both cases by finer topologies (see \cite{Heathcote},
page 255, for more details). Heathocote's arguments miss here that there are topologies
in the class $\mathfrak{Z}$ that are neither finer nor coarser than the manifold topology,
as we shall see, but our disagreement does not lie only on this ground; we believe that
the answer to the question ``what comes first, the metric or the topology'' cannot
be a definite answer in favour of the metric (see paragraph 5). There is a Lorentz metric which is ignored
by the Riemann open balls that serve as basic-open sets for the manifold topology. In addition, there are topologies different (and not finer, coarser or equal) than
the manifold topology, which incorporate the causal structure of the spacetime and
they could be considered as natural topologies for a spacetime, as well.

In recent articles (see \cite{Ordr-Ambient-Boundary} and \cite{Singularities on Amb B} and paragraph 6 here) the authors talked
about topologies in the class $\mathfrak{Z}$, in the sense that
general relativity generically leads to spacetime
 singularities where
it breaks down as a physical theory; a particular topology in $\mathfrak{Z}$ (that we have called $Z$), different than the manifold
one,
was proven to be the most natural one for this frame. It is in those papers that the authors
left as an open question a different approach to the topologisation
of spacetime: the definition of a dynamical evolution of the spacetime given
 specific causal and topological conditions (see paragraph 7). It is conjectured that the challenges or even contradictions that arise in the study
and understanding of spacetime geometry are due to the ``static'' nature of a topological
structure; indeed, there is a specific fixed topology in the background and the
topological properties arising from this topology affect the spacetime as a whole.
 This rigidity in the study of spacetime geometry might be cured if one develops a topological space
with an evolving topology, which will incorporate quantum and relativistic
frames within the spacetime and outside it (Planck time and length, singularities,
etc.). One of the aims of this critical survey is to open such a discussion
as well.

The different approaches in the study of spacetime geometry, and in the topologisation
of a spacetime in our case, are due to different cultural backgrounds;
Penrose states a similar argument in \cite{Road-to-reality}. Those who come
from QFT (quantum field theory), for example, and those from Einstein's general relativity
seem to view things in a different way
(here we should add those who come from a purely mathematical background, as
well).
Those from QFT, according to Penrose, would tend to take renormalizability
or, better, finiteness, as the primary aim of the union of relativity and
quantum theories.
Those having a relativistic background would take the
deep conceptual conflicts (determinism, causality, background independence) between the principles of quantum mechanics and those of general relativity to be the
centrally important issues that needed to be resolved,
and from whose resolution we should expect to move forward
to a new physics of the future. Those from a purely (theoretical)
mathematical background, coming straight from the Platonic world of Penrose,
would love to see a spacetime as an integrated mathematical entity,
a structure with physical properties coinciding harmonically
with the mathematical formulation.

It should be said that the description of fluctuating topologies, or topological transitions,
has become a debated topic in theoretical physics since the visionary introduction of the concept of {\it Spacetime Foam}, by John Wheeler in the Fifties~\cite{Wheeler}. In string theory these ideas have been explored in the early Nineties, among others, by Greene (see Refs.~\cite{Greene}), and innumerable have been the applications of the concept of spacetime foam to different problems (see e.g. Ref.~\cite{Scardigli}). In recent years, research lines emerged that aim to derive the concept of spacetime itself  from quantum entanglement. The seminal paper of Raamsdonk~\cite{Raamsdonk} paved the way to the more recent works of Susskind and Maldacena~\cite{Suss} -for a readable review see New Scientist~\cite{NS}- (authors who, by the way, are all building upon two fundamental, only apparently disconnected, papers written by Einstein in 1935,  the so called E.R. and E.P.R. papers~\cite{Einstein}).
On the other hand, already in a model of spacetime as simple as a lattice (see, for example, Ref.~\cite{JKS}) we see how the actual topology of spacetime can deeply affect the formulation of the fundamental structures of physical theories (in that case, the definition of the fundamental commutator of QM is deformed by the lattice structure of the underlying spacetime).

The opinions that are presented in this chapter can be considered
as opinions stemming from the family of pure mathematicians (plus a theoretical physicist) and it is expected that they will not easily drag the attention of a large number of physicists: it
is in our beliefs though that a spacetime as an integrated mathematical entity,
a spacetime studied as a topological space, would play a significant role to the search for
a theory of quantum gravity. In a few words, the methods
of general topology should be taken more seriously from those
working in QFT as well as those in general relativity, at least.
\subsection{On Name-giving and Notation.}

In the geometry of spacetime we introduce three relations: the
chronological order $\ll$, the causal order $\prec$ and the relation horismos $\rightarrow$. These
 relations can be extended to any {\em event space}
$(M,\ll,\prec,\rightarrow)$ having no metric (see \cite{Penrose-Kronheimer} and \cite{Penrose-difftopology}).

 In particular, we say that $x$ chronologically precedes an event $y$ -written
$x\ll y$- if $y$ lies inside the future null cone of $x$. $x$ {\em causally precedes} $y$ -written
$x \prec y$- if $y$ lies inside or on the future null cone of $x$. Last, but not least, $x$ is at {\em horismos}
with $y$ -written $x \rightarrow y$- if $y$ lies on the future null cone of $x$.
The order $\ll$ is irreflexive, the order $\prec$ is reflexive and the relation $\rightarrow$ is reflexive, too.

In addition, the chronological future of an event $x$ is denoted by  $I^+(x) = \{y \in M : x \ll y\}$ while its
  causal future by $J^+(x) = \{y \in M : x \prec y\}$ (with a minus instead of a plus sign, dually,  for the pasts
  in each case, respectively). The future null cone of $x$ is denoted by $\mathcal{N}^+(x)\equiv\partial J^{+}(x)= \{y \in M : x \rightarrow y\}$ and, dually, we put a minus for the null past of $x$. The chronological past and future of an event $x$ determine
  its {\em time cone}, its causal past and future its {\em causal cone} and its null past and future its {\em light cone}.

  When physicists refer to the {\em null cone} of an event $x$ they actually mean the light cone. Zeeman,
  as a working topologist, preferred to substitute the term null cone by three terms, for
  working with the interior, closure, boundary and exterior of it (see paragraph 3, of \cite{Zeeman1}).

We should now mention a few problems in name-giving that arise from when one corresponds order-theoretic
and topological notions from the classical theory of ordered sets and lattices to a spacetime manifold.
Following the construction of the {\em interval topology} (see \cite{Compendium}),
it seems natural to say that
a subset $A \subset X$ is a {\em past set} if $A = I^-(A)$ and a {\em future set} if
$A = I^+(A)$. One then would expect that the {\em future topology} $\cT^+$ is generated
by the subbase $\mathcal{S}^+ = \{X \setminus I^-(x) : x \in X\}$
and the {\em past topology} $\cT^-$ by $\mathcal{S}^- = \{X \setminus I^+(x)  : x \in X\}$.
Then, the {\em interval topology} $\cT_{in}$ on $M$ would consist of basic sets which are finite intersections
of subbasic-open sets of the past and the future topologies.

First of all, the names ``future topology'' and ``past topology'' are due to the lack of inspiration for
other names for
such topological analogues in a spacetime, but here we should have in mind that when one
considers the chronological relation and identifies $\downarrow\{x\}$ with
$I^-(x)$, then obviously $x \notin I^-(x)$. On the contrary, things follow
the pattern of the construction in \cite{Compendium} when one considers the
causal order $\prec$. Furthermore, $M \setminus I^-(x)$ will not be a future
set with $\ll$, according to the definition that a future set satisfies
$X= \uparrow X$. All these are not real problems at all, when it comes to our
target to describe particular topologies which incorporate the causal
structure of a spacetime (see the section Topologies Different than the Manifold Topology,
below, and the corresponding references in it); the problem is sort
of corresponding more appropriate names to these topologies, as well
as developing a more systematic and simplified notation. We believe that this
is not a difficult task to achieve in the near future.

One more point, regarding the appropriateness of a name; the Minkowski space in particular (and spacetimes in general)
is not up-complete, and a topology $\cT_{in}$ for a spacetime belongs actually to
 a coarser topology
than the interval topology of \cite{Compendium}. So we will treat the interval topology
of \cite{Compendium} as a special case referring to up-complete sets, and our
$\cT_{in}$ spacetime topologies belonging to a more general case where up-completeness
is not a necessary condition. It is
worth mentioning though that for the particular case of $2$-dimensional Minkowski spacetime,
$\cT_{in}$ under $\prec$ is the interval topology that one defines using \cite{Compendium}.

Finally, we would also like to highlight the distinction between
the interval topology $\cT_{in}$  from the ``interval topology'' of A.P. Alexandrov (see \cite{Penrose-difftopology}, page 29
and the succeeding section here).
$\cT_{in}$ is of a more general nature, and it can be defined via any relation, while the Alexandrov
topology is restricted to the chronological order. These two topologies are different in nature,
as well as in definition, so we propose the use of ``interval topology'' for $\cT_{in}$ exclusively,
and not for the Alexandrov topology.

\section{Topologies coarser than or equal to the manifold topology.}

In the literature, starting from the first modern singularity theorem by Penrose (see \cite{Penrose-1965})
till recent accounts on singularities such as \cite{On-Penrose-1965}, there is
no explicit mentioning of the topology of a spacetime $M$, while Riemann metric
and Riemann basic-open balls can be used whenever there is need, for example
for the proof of the Limit Curve Theorem (LCT) and the convergence of causal curves
(for a detailed exposition see \cite{Cotsakis-Singularities} and \cite{Limit-Curve-Theorems}).
In addition to the manifold topology $\mathcal{M}$, one can consider the Alexandrov topology $\mathcal{A}$ which
has basic-open sets known as ``diamonds'' and are simply the intersections
of future and past time-cones, of two distinct events respectively. This
topology incorporates the causal structure of a spacetime, but equals
the manifold topology only in the following case (see \cite{Penrose-difftopology}) and,
in other cases, it is coarser than $\mathcal{M}$.

\begin{theorem}\label{theorem strong causality}
On a spacetime $M$, the following are equivalent:
\begin{enumerate}

\item $M$ is strongly causal.

\item $\mathcal{A}$ agrees with $\mathcal{M}$.

\item $\mathcal{A}$ is Hausdorff.

\end{enumerate}

\end{theorem}

So, the main contribution of the topology $\mathcal{A}$ is a characterisation
of strong causality, as soon as $\mathcal{A}$ is Hausdorff. Adding the fact
that it incorporates the causality (in particular the chronology) of a spacetime by the construction of
open diamonds, $\mathcal{A}$ looks like a great candidate for a spacetime
topology when it is Hausdorff but, following Zeeman's arguments, its group of homeomorphisms is vast
and of no physical meaning, both in the Minkowski spacetime and in curved spacetimes.

The existence of a Lorentz metric in a spacetime is enough to make us conclude
that neither the manifold topology $\mathcal{M}$ nor the Alexandrov topology $\mathcal{A}$
``in its best'', that is when Theorem \ref{theorem strong causality} is satisfied, can
fully describe a spacetime topologically. The manifold topology is a natural topology
for a manifold, but not such a natural one for a spacetime manifold!

\section{The class $\mathfrak{Z}$ of Zeeman-G\"obel topologies.}

The class $\mathfrak{Z}$ of Zeeman topologies
on a spacetime manifold $M$ consists of topologies
which have the property that they induce the $1$-dimensional manifold topology on every time
axis and the $3$-dimensional manifold topology on every space
axes. This class was first introduced in \cite{Zeeman1}, in the
special case of Minkowski spacetime, and it was generalised in
\cite{gobel} for any curved spacetime. In particular, paper
\cite{Zeeman1} is the natural continuation of \cite{Zeeman2},
where Zeeman proved that causality in Minkowski spacetime implies
the Lorentz group. He then showed that the group of all homeomorphisms
of the finest topology in $\mathfrak{Z}$, which is coarser than the discrete
topology, is generated by the inhomogeneous Lorentz group and dilatations.
In addition, unlike the topology of $\mathbb{R}^4$, this fine
topology $F$ is not locally homogeneous and the light
cone through any point can be deduced by $F$. There is also
a quite interesting lemma; the topology on a light ray induced
from $F$ is discrete. Discreteness of light, according to G\"obel,
describes well its physical behaviour: there is no geometric
information along a light ray. Here one should not confuse topological
discreteness (every set is open) with discreteness in the sense
of (finite or infinite) countability. Apart from the group
of homeomorphisms of $M$ under $F$ and its physical interpretation,
the topological boundary of the null cone has the maximum number
of open sets: there is definitely a connection here with the maximum
speed, that of light.

Zeeman mentioned three other alternative topologies in
$\mathfrak{Z}$ different than $F$,
that we will consider in section 4, as well as their
analogues for curved spacetimes.

G\"obel found that the analogue of $F$ in a curved spacetime
has the property that the group of all homeomorphisms under
this topology is isomorphic to the group
of all homothetic transformations. In a few
words, under the relativistic analogue of $F$, a homeomorphism is an isometry.

A problem, that was noticed first by Zeeman himself, is that $F$
is technically difficult, as it
does not admit a countable base and so it is not the best tool
for a working physicist. This was one of the arguments of
the authors of \cite{Hawking-Topology} and \cite{Low_path} as well,
but we object that this is not an attractive reason for avoiding
a topology which is much more natural in a spacetime from
the manifold topology. Natural in the sense that it incorporates the differential,
causal and conformal structures and the group of homeomorphisms
of the spacetime is not vast and it has physical meaning. So,
the argument that $F$ has ``too many
open sets'' and does not admit a countable base should
be reconsidered.
Since we are dealing with both the Lorentz metric as well
as the Riemann metric in a spacetime manifold, a natural
topology which will describe the properties of the spacetime
should be compatible with every possible structure which
is defined on the spacetime. $F$ is such a topology.

For some reason the supporters
of the manifold topology, like Heathcote,
believed that all the topologies in $\mathfrak{Z}$ are strictly
finer than $\mathcal{M}$, but actually this is not true. Three  alternative topologies that Zeeman introduces
in \cite{Zeeman1} are linked in their construction to
topologies that we mention in the next paragraph, each of
which belongs to the class $\mathfrak{Z}$ but is incomparable to $\mathcal{M}$ (see \cite{Limit-Curve-Theorems}, \cite{Order-Light-Cone}
and \cite{On-Two-Zeeman-Topologies}). G\"{o}bel (\cite{gobel} page 297, (C)) actually
states that there are other topologies in $\mathfrak{Z}$, but without a clear reference that there are topologies
that are not necessarily finer or coarser or equal to the manifold topology. This is important, since the criticism against
the class $\mathfrak{Z}$ bases many of its arguments against the term {\em finer} topology.
Let us now look at a sample of topologies in $\mathfrak{Z}$, which
are not finer than the manifold topology $\mathcal{M}$.

\section{Topologies different than the manifold topology.}

In \cite{LimitCurve} we remark that the Path Topology
$\mathcal{P}$ of Hawking-King-McCarthy (see \cite{Hawking-Topology}) is
the general relativistic analogue of the topology introduced in Example 1
of \cite{Zeeman1} (page 169). Low showed in \cite{Low_path} that under this
topology $\mathcal{P}$ (that we name $Z^T$ for consistency of notation) the Limit Curve Theorem (LCT) fails to hold.
In \cite{LimitCurve} we introduced three (among others) more topologies that
the LCT fails to hold, all incorporating the differential, causal and conformal structure
of the spacetime manifold. In particular, we stated the following theorem.

\begin{theorem}\label{2}
There are three distinct topologies in a spacetime manifold which admit a countable basis, they
incorporate the causal and conformal structures and the LCT fails with each one of them respectively.
These are the interval topologies
$T_{in}^{\rightarrow}$, $T_{in}^{\le}$ and $T_{in}^{\ll^=}$, which are all in the class
$\mathfrak{Z}$.
\end{theorem}

All these topologies are not finer (neither equal nor coarser) than the manifold topology
and singularity theorems, under each one of them respectively, cannot be formed in the way
that are described via the manifold topology. These three topologies, together
with the manifold topology, give the intersection topologies $Z$, $Z^T$, $Z^S$,
which are finer than the manifold topology, where $Z$ is coarser than the Fine
topology $F$ and $Z^T$ (the Path topology of \cite{Hawking-Topology}) and $Z^S$
are incomparable to $F$.

 Low, in \cite{Low_path}, supports in his conclusion
that LCT failing in the $\mathcal{P}$ (which also fails in the extra
five topologies that we suggest in Theorem \ref{2}) makes the manifold topology remaining both
technically easier to work with and fruitful. We have some objections. All the six topologies
of \cite{LimitCurve} and in particular those in Theorem \ref{2} are technically easy to work with (they all have a countable base
of open sets) and they are fruitful, as they belong to $\mathfrak{Z}$ and are all
behaving like order topologies, in the sense that they satisfy the orderability
problem (or weaker versions of it, referring to non-linear orders; see \cite{Good-Papadopoulos},
\cite{Orderability-Theorem}, \cite{Nestsandtheirrole} and \cite{OnProperties}). Each one of them is induced either from the causal or chronological
orders or from (the irreflexive) horismos, with the exception of $Z^S$ which is induced by a particular
spacelike non-causal order that we describe in \cite{On-Two-Zeeman-Topologies}.
More specifically,
$Z$, $Z^T$ and $Z^S$ have open sets bounded by Riemann open balls centered at an event
$x$, intersected with the timecone union spacecone of $x$ in the case of $Z$, the timecone of $x$ in the case
of $Z^T$ and
the spacecone of $x$ in the case of $Z^S$, respectively. The rest three topologies have unbounded
open sets which are timecone union spacecone in the case of $T_{in}^{\rightarrow}$, timecone in
the case of $T_{in}^{\le}$ and, spacecone in the case of $T_{in}^{\ll^=}$, respectively,
at an event $x$. For a more detailed treatment we refer to \cite{LimitCurve}.

On the other hand, the manifold topology misses
the Lorentz metric and so the causal structure of the spacetime as well, so we conclude
the following.

\begin{corollary}\label{3}
The manifold topology $\mathcal{M}$, on a spacetime $M$, is based on the Riemann metric and is sufficient for describing
spacetime singularities, but does not incorporate the Lorentz metric, while each of the topologies in Theorem \ref{2} fail
to describe singularities that appear under $\mathcal{M}$, but incorporate the Lorentz metric.
\end{corollary}

\begin{corollary}\label{4}
{{\bf The Fine Topology $F$ is the best possible candidate
for a spacetime}} $M$, as it is strictly finer than $\mathcal{M}$, strictly
coarser than the discrete topology and, simultaneously,
finer than the topologies introduced in Theorem \ref{2}. In addition,
the group of homeomorphisms of $M$ under $F$ is isomorphic to the Lorentz group
and dilatations, in the case of special relativity, and to the group of
homothetic symmetries in the case of general relativity, while under the
manifold topology the group of homeomorphisms of $M$ is vast and of no
physical significance. Last, but not least, the LCT holds under $F$, while
it might fail in coarser topologies to $F$.
\end{corollary}

The discussion about $F$ would be incomplete, if we did not mention the
comment of G\"obel about $F$ in \cite{gobel}, pages 290-291: unphysical world
lines, like ``bad trips'' are avoided if one interprets continuity of
worldlines with respect to $F$. Under $F$ basic assumptions for a kinetic
theory in general relativity are satisfied and one can incorporate 
electromagnetic fields into such a result, if one allows $F$ to depend on
a gravitational field as well as on the Maxwell field (and derive corresponding
results on orbits of freely falling test particles for charged particles).

\section{In the beginning was the metric...or the topology?}

This is a more important question as it seems to be. Speaking
about spacetime manifolds as mathematical objects, it is vital
that a natural topology will incorporate all the mathematical structures
appearing in the manifold, including the Lorentz metric as well as the Riemann
metric. In this sense (a ``Platonic mathematical'' sense in the view of Penrose,
which is projected to the physical world \cite{Road-to-reality})
the manifold topology is not a natural topology in a spacetime manifold,
even if it is defined via the Riemann metric. The metric tensor
field, the affine connection and the conformal structure,
the differentiable manifold with its topology, are all important
constituents of the spacetime manifold, but what about the Lorentz
metric and the structure of the null cone? Having mentioned this, we
believe that ``in the beginning was the topology'', in a spacetime manifold. A topology
like $F$, where the group of homeomorphisms of $M$ under
$F$ has a physical meaning and which incorporates all the metric
structures in the manifold.

\section{Ambient cosmology: a failure due to a topological misconception.}

In  \cite{Ordr-Ambient-Boundary} we described the motivation for a $5$-dimensional
``ambient space'', where our $4$-dimensional spacetime is its conformally related ambient boundary at
infinity, by linking it to the singularity problem in general relativity.
In cosmology the infinities that are inherent in the spacetime metric according to the singularity
theorems indicate the necessity of a conformal geometry of metrics to absorb them,
not a breakdown of general relativity. The construction of this model in ambient
cosmology can be found in \cite{Cotsakis1}, \cite{Cotsakis2}, \cite{Cotsakis3}
and \cite{Cotsakis4}, where the authors started from the construction of the metric,
leaving the topological problem at the end. As we observed in \cite{Singularities on Amb B},
it is the topology succeeding (and, unfortunately, not preceding or at least
being constructed simultaneously with) the metric that showed a failure in the
construction and in results concerning the convergence of causal curves; it is the Path topology $\mathcal{P} = Z^T$ or $Z^S$ or $Z$ where the LCT
fails and not in $F$. Furthermore, why should one bother to add an extra dimension
while a $4$-dimensional spacetime under a topology like $\mathcal{P}$ has already the properties
of the ambient boundary, and while the structure of the ambient boundary is totally unknown to us (we lack
knowledge even for
basic results on causality: see, for example, \cite{Order-Light-Cone} for an important correction
on \cite{Ordr-Ambient-Boundary}). Here we should also mention that a finer topology than the 
manifold one will contain manifold-open sets; this does not guarantee that the LCT holds. For example,
$\mathcal{P}$ is finer than $\mathcal{M}$ and, simultaneously, LCT fails under $\mathcal{P}$.

\section{Towards an evolving topology and a Quantum Theory of Gravity.}

If the main problem for a working physicist is that $F$ is not an easy
topology to work with, due to the lack of a countable basis of open sets,
or if topologies like $\mathcal{M}$ and those topologies mentioned in paragraph 4
are missing something important from the spacetime structure, then we
believe that there is something deeper behind all this and this
certainly is of a topological nature. We have already expressed in \cite{Order-Light-Cone}
an idea of an evolving topology with respect to the class $\mathfrak{Z}$, so that
different topologies of this class are assigned to each stage of
the evolution as well as where the spacetime itself is subjected to singularities. It could be, for example, that the interval topology from horismos
$\rightarrow$ (see \cite{Order-Light-Cone}) could give a sufficient description to the transition from/to the Planck time and objects like black holes, while
other topologies (where the LCT theorem holds for example) could explain the
phase transition from locality to non-locality. Topologies like $Z^T$ are
linked to a discrete space while $Z^S$ to a discrete time, while $Z$ to
a discrete light (these are actually remarks of Zeeman in \cite{Zeeman1}, for
their special relativistic analogues). By evolution we do not necessarily mean (and this is not our desire
at all) to consider kinematically that the spacetimes of our interest are foliated
manifolds where leaves of foliation have open sets which vary over time. The question
is different: how does a spacetime manifold appear from a functional space? An
answer to such a question which refers to the transition from nonlocality
to locality seems to need a richer topological background; a backgrould that the class
$\mathfrak{Z}$ could possibly provide. Possible tools can be also derived from articles
like \cite{Nanda}, \cite{Agarwal}.

\section{The need to take methods of general topology more seriously.}

We believe that the concerns against a ``finer'' topology, as
expressed in \cite{Heathcote}, are reasonable. Reasonable are
similar concerns expressed in \cite{Low_path}; when we restrict
ourselves to the validity of general relativity. The problem
is that eventhough the manifold topology $\mathcal{M}$ has
somehow worked nicely in the last century or so, it is problematic in describing
fully properties of a spacetime in a sufficient way; it lacks important
information, as we have seen in the previous paragraphs. $F$ is a finer topology which resolves, at
least in a mathematical way,
all such issues, and -at the moment- there is no other candidate topology to compete with.

Criticism (in oral communication with physicists) against $F$, and against topologies like those mentioned
in paragraph 4, highlight that
there is a value of considering these alternative topologies in
$\mathfrak{Z}$ since
they may, for example, lead to a new physical theory; or they may
allow one to extract new, physically interesting, predictions from
the old theory. But it seems, according to the critics, that there is a point of
diminishing returns; that, eventually, further treatment of these
topologies, in the abstract, can no longer be justified. At some point,
there is a burden to extract from these topologies, some concrete
result of genuine physical interest; no such
result is in sight and, therefore, that we have reached that point.

The problem of such a criticism is that the main points of \cite{Zeeman1}
and \cite{gobel} have not been understood, and this is quite disappointing.
There is a prejudice against general topology; only the reference to it
is enough to discourage working mathematical physicists and theoretical
physicists to read carefully a related article. The labyrinth that we seem
to be when talking about string theory and quantum theory of gravity, for example,
is not only related to the need for an extra physical input, but for
an extra mathematical input as well.
 The authors wish that this
Chapter contributes to the reopening of a discussion in this serious and
fascinating subject.

\section*{Acknowledgements.}  The co-author K.P. wishes to
thank Robert Low for his remarks on the Interval Topology in \cite{LimitCurve}, some of
which we incorporate in the introductory section here, as well as
Spiros Cotsakis for introducing him \cite{Zeeman1} and related literature. He also wishes to thank to Nikolaos Kalogeropoulos
for discussions on quantum theory of gravity and B.K. Papadopoulos for having taught him lattice
topologies. Last, the authors would like to thank Ljubisa Kocinac and Hemen Dutta
for their important remarks towards the improvement of the text.


\begin{thebibliography}{99}

\bibitem{Ordr-Ambient-Boundary} I. Antoniadis, S. Cotsakis and K. Papadopoulos,
The Causal Order on the Ambient Boundary, Mod. Phys. Lett. A, Vol 31, Issue 20, 2016.

\bibitem{Cotsakis1} I. Antoniadis, S. Cotsakis, Ambient cosmology
and spacetime singularities, Eur. Phys. J.C. 75:35 (2015) 1-12.

\bibitem{Cotsakis2} I. Antoniadis, S. Cotsakis, Topology of the ambient boundary and the convergence of causal curves, Mod. Phys. Lett. A, Vol. 30, No. 30 (2015) 1550161.

\bibitem{Cotsakis3} I. Antoniadis, S. Cotsakis, The large-scale structure of the ambient boundary,
to appear in MG14 Proceedings; arXiv:1512.0916.

\bibitem {Greene} P.S. Aspinwall, B.R.Greene, D.R. Morrison, Multiple Mirror Manifolds and Topology Change in String Theory, Phys.Lett. B303 (1993) 249-259.\\
 B.R. Greene, K. Schalm, G. Shiu, Dynamical Topology Change in M Theory, J.Math.Phys. 42, 3171-3187 (2001).

\bibitem{Cotsakis-Singularities} S. Cotsakis,  Talking about Singularities, The Eleventh Marcel Grossmann Meeting: On Recent Developments in Theoretical and Experimental General Relativity, Gravitation and Relativistic Field Theories (In 3 Volumes), pages 758-777, 2008.

\bibitem{Cotsakis4} S. Cotsakis, Cosmological Singularities,
Springer LNP Proceedings of the First Aegean Summer School of Cosmology, Samos, September 21-29, 2001.

\bibitem{Einstein} A.~Einstein and N.~Rosen, The Particle Problem in the General Theory of Relativity, Phys.\ Rev.\  {\bf 48} (1935) 73.\\
A.~Einstein, B.~Podolsky and N.~Rosen, Can quantum mechanical description of physical reality be considered complete?, Phys.\ Rev.\  {\bf 47} (1935) 777.

\bibitem{Compendium}  G. Gierz, K.H. Hofmann, K. Keimel, J.D. Lawson, M.W. Mislove and D.S. Scott, A compendium of continuous lattices, Springer-Verlag, 1980 .

\bibitem{gobel} Gobel, Zeeman Topologies on Space-Times of General Relativity Theory, Comm. Math. Phys. 46, 289-307 (1976).

\bibitem{Good-Papadopoulos} Good Chris, Papadopoulos Kyriakos, A topological characterization of ordinals: van Dalen and Wattel revisited, Topology Appl. 159 (2012), 1565-1572

\bibitem{Hawking-Topology} S.W. Hawking, A.R. King, P. J. and McCarthy, A new topology for curved space–time which incorporates the causal, differential, and conformal structures. Journal of Mathematical Physics, 17 (2). pp. 174-181, 1976.

\bibitem{Heathcote} A. Heathcote, Zeeman G\"{o}bel Topologies, Brit. J. Phil. Sci. 39 (1988), 247-261.


\bibitem{JKS} P.~Jizba, H.~Kleinert and F.~Scardigli,  Uncertainty Relation on World Crystal and its Applications to Micro Black Holes, Phys.\ Rev.\ D {\bf 81} (2010) 084030,
  [arXiv:0912.2253 [hep-th]].

\bibitem{Low_path} R.J. Low, Spaces of paths and the path topology, Journal of Mathematical
Physics, 57, 092503 (2016).

\bibitem{Penrose-Kronheimer} E.H. Kronheimer and R. Penrose, On the structure of causal spaces,
Proc. Camb. Phil. Soc. (1967), 63, 481.

\bibitem{Suss} J.~Maldacena and L.~Susskind, Cool horizons for entangled black holes,
  Fortsch.\ Phys.\  {\bf 61} (2013) 781, [arXiv:1306.0533 [hep-th]].\\
	L.~Susskind, {\sl Dear Qubitzers, GR=QM}, arXiv:1708.03040 [hep-th].

\bibitem{Limit-Curve-Theorems} E. Minguzzi, Limit curve theorems in Lorentzian geometry, J. Math. Phys. 49, 092501 (2008).

\bibitem{NS} New Scientist Cover Story,  Entangled Universe, 7 November 2015.

\bibitem{Orderability-Theorem} K. Papadopoulos, On the Orderability Problem and the Interval Topology, Chapter in the Volume ``Topics in Mathematical Analysis and Applications'', in the Optimization and Its Applications Springer Series, T. Rassias and L. Toth Eds, Springer Verlag, 2014.

\bibitem{OnProperties} K. Papadopoulos,  On Properties of Nests: Some Answers and Questions, Questions and Answers in General Topology, Vol. 33, No. 2 (2015).

\bibitem{Nestsandtheirrole} K. Papadopoulos,   Nests, and their role in the Orderability Problem, Mathematical Analysis, Approximation Theory and their Applications, pp 517-533, Th. M. Rassias and V. Gupta Eds, Springer (2016).

\bibitem{Singularities on Amb B} K. Papadopoulos, On the possibility of singularities on the ambient boundary, International
Journal of Geometric Methods in Modern Physics, Vol. 14, No. 10, 2017.

\bibitem{Order-Light-Cone} K. Papadopoulos, S. Acharjee and B.K. Papadopoulos, The Order On the Light Cone and Its Induced Topology, International Journal of Geometric Methods in Modern Physics, Vol. 15, No. 05, 1850069 (2018).

\bibitem{On-Two-Zeeman-Topologies} K. Papadopoulos and B.K. Papadopoulos, On Two Topologies that were suggested by Zeeman,
accepted in Mathematical Methods in the Applied Sciences (DOI: 10.1002/mma.5238).

\bibitem{LimitCurve} K. Papadopoulos and B.K. Papadopoulos,  Spacetime Singularities vs. Topologies of Zeeman-G\"obel Class,
 arXiv: 1712.03270.

\bibitem{Penrose-1965} Penrose R, Gravitational collapse and space-time singularities. Phys. Rev. Lett. 14, 57, 1965.

\bibitem{Penrose-difftopology} R. Penrose, Techniques of Differential Topology in Relativity,
CBMS-NSF Regional Conference Series in Applied Mathematics, 1972.

\bibitem{Road-to-reality} R. Penrose,  The Road to Reality: a complete guide to the laws of the universe, Vintage Books, 2007 edition.

\bibitem{Scardigli} F. Scardigli, Black hole entropy: A Space-time foam approach,
  Class.\ Quant.\ Grav.\  {\bf 14} (1997) 1781 [gr-qc/9706030].

\bibitem{On-Penrose-1965} J.M.M. Senovilla, and D. Garfinkle, The 1965 Penrose singularity theorem, Classical and Quantum Gravity, Volume 32, Number 12, 2015.

\bibitem{Raamsdonk} M.~Van Raamsdonk, Building up spacetime with quantum entanglement,
  Gen.\ Rel.\ Grav.\  {\bf 42} (2010) 2323
   [Int.\ J.\ Mod.\ Phys.\ D {\bf 19} (2010) 2429]
  [arXiv:1005.3035 [hep-th]].

\bibitem{Wheeler} J.A. Wheeler, Geons, Phys.\ Rev.\  {\bf 97} (1955) 511.

\bibitem{Zeeman2} E.C. Zeeman,  Causality implies the Lorentz group,
J. Math. Phys. 5 (1964), 490-493.

\bibitem{Zeeman1} E.C. Zeeman, The Topology of Minkowski Space,
Topology, Vol. 6, 161-170(1967).


\end{thebibliography}
\end{document}